\documentclass[conference]{IEEEtran}
\usepackage{cite}
\usepackage{graphicx}
\usepackage{psfrag}
\usepackage{url}
\usepackage{amsmath}
\usepackage{array}
\usepackage{amssymb}
\usepackage{mathtools}
\usepackage{amsfonts}
\usepackage{graphicx}
\usepackage{epstopdf}
\usepackage{algorithm}
\usepackage{algpseudocode}
\newtheorem{proposition}{Proposition}

\usepackage{setspace}
\usepackage{amscd}
\usepackage{mathrsfs}
\usepackage{epsfig}
\usepackage{color}
\usepackage{textcomp}
\usepackage{multirow}
\usepackage{caption}
\usepackage{subcaption}
\usepackage{threeparttable}
\title{Space-Time Coded Spatial Modulated Physical Layer Network Coding for Two-Way Relaying}
\begin{document}

\author { %{K.G. Unnikrishnan and B. Sundar Rajan, ˜\IEEEmembership{˜Fellow ˜IEEE } }
\authorblockN{K.G. Unnikrishnan and B. Sundar Rajan}
\authorblockA{Dept. of ECE, IISc, Bangalore 560012, India, Email:{$\lbrace$unnikrishnankg89, bsrajan$\rbrace$}@ece.iisc.ernet.in
}
}

\maketitle
%%%%%%%%
\begin{abstract}
Using the spatial modulation approach, where only one transmit antenna is active at a time, we propose two transmission schemes for two-way relay channel using physical layer network coding with space time coding using Coordinate Interleaved Orthogonal Designs (CIOD's). It is shown that using two uncorrelated transmit antennas at the nodes, but using only one RF transmit chain and space-time coding across these antennas can give a better performance without using any extra resources and without increasing the hardware implementation cost and complexity. In the first transmission scheme, two antennas are used only at the relay, Adaptive Network Coding (ANC) is employed at the relay and the relay transmits a CIOD Space Time Block Code (STBC). This gives a better performance compared to an existing ANC scheme for two-way relay channel which uses one antenna each at all the three nodes. It is shown that for this scheme at high SNR the average end-to-end symbol error probability (SEP) is upper bounded by twice the SEP of a point-to-point fading channel. In the second transmission scheme, two transmit antennas are used at all the three nodes, CIOD STBC's are transmitted in multiple access and broadcast phases. This scheme provides a diversity order of two for the average end-to-end SEP with an increased decoding complexity of $\mathcal{O}(M^3)$ for an arbitrary signal set and $\mathcal{O}(M^2\sqrt{M})$ for square QAM signal set.
\\
\end{abstract}

\begin{IEEEkeywords} 
Denoise-and-forward, wireless two-way relaying, physical layer network coding, coordinate interleaved orthogonal designs, spatial modulation.
\end{IEEEkeywords}

\section{Introduction} 
\begin{figure}[htbp]
\centering
\includegraphics[scale=0.5]{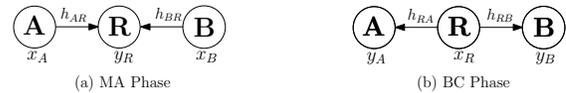}
\caption{The two-way relay channel}
\label{relay_channel}
\end{figure}

Wireless two-way relaying scenario in Fig. \ref{relay_channel} is considered where bidirectional data transfer takes place between nodes A and B with the help of a relay R. The node A(B) cannot directly communicate with the other end node B(A).  It is assumed that all the three nodes are operating in half duplex, i.e. the nodes cannot transmit and receive simultaneously in the same frequency band. Physical Layer Network Coding (PLNC) introduced in \cite{ZhLiLa}, is a new paradigm for wireless networks where the relay in the network jointly decodes the transmitted messages and broadcasts a function of the decoded messages. This provides a significant increase in the achievable rates in some networking scenario. PLNC exploited the Multiple Access Interference (MAI) occurring at the relay in two-way relay channels, so that communication between the end nodes requires two phase protocol. The relaying protocol consists of two phases, the \textit{Multiple Access} (MA) phase where the nodes A and B simultaneously transmit to the relay R, Fig. \ref{relay_channel}(a) and \textit{Broadcast} (BC) phase where relay R transmits a function of the received  signals during MA phase to the nodes A and B, Fig \ref{relay_channel}(b). At the relay R, network coding is employed so that A(B) can decode B's(A's) message, given that A(B) knows its own message.

Information theoretic studies about the achievable rates using physical layer network coding  is reported in \cite{NaGa},\cite{HeNa},\cite{PoYo1}, \cite{PoYo2}. Compute and forward strategy in \cite{NaGa}, exploits the multiple access interference to increase the achievable rates between the users in the network. Encoder maps messages from a finite field to lattice and the decoder recovers the lattice points from the function of the transmitted lattice points and maps back a function over the finite field. In \cite{HeNa}, in the compute and forward paradigm, the achievable rates can be increased than those achievable by fixed functions by using a multilevel coding scheme based on the channel gains. Denoise and forward (DNF) protocol proposed in \cite{PoYo1}, \cite{PoYo2} further improves the achievable rates by combining symbols in the multiple access channel and removing noise from the combined symbols before broadcasting. 

A significant improvement in the end-to-end throughput for two-way relay channel using Adaptive Network Coding (ANC) schemes based on DNF protocol is proposed in \cite{APT1}, \cite{NVR}. The impact of MAI is minimized by adaptively changing the network coding maps according to the channel conditions. In \cite{APT1}, the adaptive network coding maps were obtained using computer search algorithms and in \cite{NVR}, adaptive network coding maps were obtained analytically using Latin squares approach.

Spatial Modulation (SM) introduced in \cite{SM}, is a recent transmission technique for multiple antenna systems mitigating the Inter Channel Interference (ICI). Practical multi antenna systems require multiple Radio Frequency (RF) chains at the transmitter and receiver, increasing the associated cost, hardware complexity and inter antenna synchronization procedures. In SM only one transmit antenna is active at a time, thus avoiding the above issues and the data is conveyed both through the symbol chosen from constellation and unique transmit antenna number chosen from set of transmit antennas. For the same spectral efficiency, SM results in lesser decoding complexity compared to Vertical Bell Labs Layered Space Time (V-BLAST) scheme. In \cite{GSM}, Generalized Spatial Modulation (GSM) uses an optimum number of RF transmit chains and activates more than one transmit antenna to achieve better throughput than spatial multiplexing. Diversity gains are exploited for SM schemes in \cite{STBCSM}, \cite{CISTSM} using space time coding. It was shown in \cite{CISTSM}, that using one RF transmit chain in SM, diversity order of two is achievable using Coordinate Interleaved Orthogonal Designs(CIOD) in space time coding. Hence diversity gain can be achieved without increasing the hardware cost and complexity. In \cite{SMTWC}, spatial modulation for two-way network coded channel have been studied and performance analysis is carried out when the end nodes use space shift keying (SSK) \cite{SSK}. 

In this paper, for wireless two-way relaying scenario, we consider all the nodes have only one RF transmit chain similar to spatial modulation technique, thereby using the same resources as in \cite{APT1}, \cite{NVR}, \cite{VvR_DSTC}. We propose two transmission schemes using physical layer network coding with space time coding with CIOD's for two scenarios, Scheme-I: when only the relay has two transmit antennas and other nodes have one antenna and Scheme-II: when all the three nodes have two transmit antennas.

The contributions of this paper may be summarized as follows:
\begin{itemize}
\item For the proposed Scheme-I, coding gain can be achieved compared to the schemes in \cite{APT1}, \cite{NVR}. This coding gain is due to the reduction of BC phase errors by using space time coding. 
\item For the proposed Scheme-II, a diversity order of two can be achieved by physical layer network coding with space time coding at all nodes using CIOD's compared to the schemes in \cite{APT1}, \cite{NVR} and \cite{VvR_DSTC} which have diversity order one. But this advantage comes at a cost of increased decoding complexity at R.
\end{itemize}

Hence just by adding an extra uncorrelated antenna at the nodes for transmission in the existing schemes \cite{APT1}, \cite{NVR}, \cite{VvR_DSTC}, without using a second RF chain, and space-time coding over these antennas can achieve coding and diversity gains. Hence the average end-to-end symbol error probability can be reduced significantly without increasing the hardware cost and complexity for implementation. 

The organization of rest of the paper is as follows: Section \ref{Preliminaries} gives a brief  description of the preliminaries used in the paper. The two proposed schemes, Scheme-I and Scheme-II are explained in Sections \ref{APLNC_SM} and \ref{PLNC_SM} respectively. Simulation results are provided in Section \ref{Sim_Results}.

\textbf{\textit{Notations:}} Throughout, bold, lowercase letters are used to denote vectors and bold, uppercase letters are used to denote matrices. For a complex variable $x$, $x_I$ and $x_Q$ denote the real and imaginary part of $x$, respectively. The complex number $\sqrt{-1}$ is denoted by $j$. $\mathbb{R}$ and $\mathbb{C}$ denote the set of real numbers and the set of complex numbers respectively. Let $span(\mathbf{x_1,x_2,...,x_n})$ denote the vector subspace over $\mathbb{C}$ spanned by the complex vectors $\mathbf{x_1,x_2,...,x_n}$. For a matrix $\bold{A}$, $\bold{A}^T$ and $\bold{A}^{H}$ denotes transpose and Hermitian transpose of the matrix $\bold{A}$ respectively. For a vector subspace $V$ of a vector space over $\mathbb{C}$, $V^{\dagger}$ denotes the vector subspace $\{\mathbf{x : x^Tv}=0,\forall \mathbf{v} \in V \}$. $\mathcal{CN}(0,\sigma^2)$ denotes a circularly symmetric complex Gaussian random variable with zero mean and variance equal to $\sigma^2$. Throughout, by SNR we mean $\frac{E_s}{\sigma^2}$, where $E_s$ is the average transmission energy of all the nodes and $\sigma^2$ is the variance of the additive noises. $||.||_2$ represents the Euclidean vector norm. The operation $\otimes$ denote the Kronecker product function. $\bold{I}_N$ and $\bold{O}_N$ represents the $N \times N$ identity matrix and null matrix respectively.

\section{Preliminaries}
\label{Preliminaries}
\subsection{Coordinate Interleaved Orthogonal Designs (CIOD)} CIOD's are single complex symbol maximum likelihood decodable linear space time block codes which can offer full rate and full diversity \cite{Khan_BSR}. Diversity gain in CIOD is achieved by sending in-phase and quadrature components of a symbol through different antennas. Let $x_1 = x_{1I} + j x_{1Q}$ and $x_2 = x_{2I} + j x_{2Q}$, be the two complex symbols to be transmitted using coordinate interleaving. Then the CIOD STBC for two transmit antenna is given by
\begin{equation}
\bold{X}= \left[ \begin{array}{cc} \tilde{x}_1 & 0 \\ 0 & \tilde{x}_2 \\ \end{array} \right]
\label{CIOD}
\end{equation}
where the row and column indices denotes the antenna and time indices respectively. The transmitted symbols are $\tilde{x}_1=x_{1I}+jx_{2Q}$ and $\tilde{x}_2=x_{2I}+jx_{1Q}$. From (\ref{CIOD}), it is clear that for each channel use only one transmit antenna is active. Hence only one RF transmit chain is required for the transmission of CIOD STBC's. The operation required at the transmitter is to decouple the in-phase and quadrature components of the symbol and combine it with that of the other symbol and to switch the transmit RF chain to the required transmit antenna in every channel use.

CIOD STBC's will give full diversity when the difference in in-phase and quadrature components is non-zero, i.e. when $x_1$ and $x_2$ are taken from the signal set which is constructed to have a non-zero Coordinate Product Distance (CPD) \cite{Khan_BSR}. Further it was shown that for the CIOD STBC's constructed over square-QAM signal sets, the coding gain will be maximized by rotating the constellation by an angle $\theta=\frac{\arctan(2)}{2}$ radians and the maximum coding gain is given by $\frac{4d^2}{\sqrt{5}}$, where $d$ is the minimum distance between any two points in the square QAM signal set having unit energy. 

\subsection{Notion of Singular Fade Subspaces for Two-Way Relaying Scenario}
\label{SFC}
Consider two-way relaying scenario with two transmit antennas each at nodes A and B and one antenna at the relay. Let $\bold{h_{AR}} = [h_{A_1R} \ h_{A_2R}]$ and $\bold{h_{BR}} = [h_{B_1R} \ h_{B_2R}]$ be the fade coefficients associated with A-R and B-R links respectively, where $h_{A_iR}$ and $h_{B_iR}$ represents the fade coefficient from the $i$th antenna of node A and B respectively to R. Let $\mathbf{X_A}$ and $\mathbf{X_B}$ be $2\times T$ space time codeword matrices of the space time code $\mathcal{C}$, sent by nodes A and B respectively, where $T$ is the number of time slots for which the nodes transmit. The effective constellation, $\mathcal{S}_R(\mathbf{h_{AR}},\mathbf{h_{BR}})$ seen at R is given by, 
\[ \mathcal{S}_R(\mathbf{h_{AR}},\mathbf{h_{BR}}) =\{ \mathbf{h_{AR} X_{A}} + \mathbf{h_{BR} X_{B}} \ | \ \mathbf{X_A},\mathbf{X_B} \in \mathcal{C}  \}.\]
Let $d_{min}(\mathbf{h_{AR}},\mathbf{h_{BR}})$ be the minimum distance between the points in the effective constellation $\mathcal{S}_R(\mathbf{h_{AR}},\mathbf{h_{BR}})$, i.e.,
{ \footnotesize
\begin{align}
d_{min}(\mathbf{h_{AR}},\mathbf{h_{BR}}) = \mathop{\min_{\mathbf{X_A},\mathbf{X_B}, \mathbf{X_A'},\mathbf{X_B'} \in \mathcal{C} }}_{(\mathbf{X_A},\mathbf{X_B}) \neq (\mathbf{X_A'},\mathbf{X_B'})} ||\mathbf{h_{AR}}\mathbf{\Delta X_A} + \mathbf{h_{BR}}\mathbf{\Delta X_B} || 
\label{Singspace}
\end{align} }
where $\mathbf{\Delta X_A} = \mathbf{X_A} - \mathbf{X_A'}$ and $\mathbf{\Delta X_B} = \mathbf{X_B} - \mathbf{X_B'}$. From (\ref{Singspace}), it is clear that there exists values for $\mathbf{h_{AR}}$ and $\mathbf{h_{BR}}$ such that $d_{min}(\mathbf{h_{AR}},\mathbf{h_{BR}}) = 0 $. The minimum distance in the effective constellation will be zero when the vector $[\mathbf{h_{AR} \ h_{BR}}]$ fall in a vector subspace of the form { \scriptsize $\left( span \left( \left[ \begin{array}{c}
\mathbf{\Delta X_A}\\ \mathbf{\Delta X_B} \end{array} \right]  \right) \right)^{\dagger}$ }, known as the \textit{singular fade subspaces} \cite{VvR_DSTC}. The probability that the vector $[\mathbf{h_{AR} \ h_{BR}}]$ belongs to a singular fade subspace is zero, but when the vector $[\mathbf{h_{AR} \ h_{BR}}]$ falls close to a singular fade subspace,	  $d_{min}(\mathbf{h_{AR}},\mathbf{h_{BR}})$ is greatly reduced, a phenomenon known as \textit{distance shortening} \cite{NVR}, \cite{VvR_DSTC}. 

Let $\mathcal{V}$ be a singular fade subspace. A map $\mathcal{M}^{\mathcal{V}}$ is said to remove the singular fade subspace $\mathcal{V}$, when $\mathcal{M}^{\mathcal{V}}(\mathbf{X_A},\mathbf{X_B}) = \mathcal{M}^{\mathcal{V}}(\mathbf{X_A'},\mathbf{X_B'})$, $\forall \mathbf{X_A,X_B,X_A',X_B'}$ such that $\mathcal{V} = $  { \scriptsize   $\left( span \left( \left[ \begin{array}{c}
\mathbf{X_A-X_A'} \\ \mathbf{X_B-X_B'}
\end{array} \right]  \right)  \right)^{\dagger} $ }. The harmful effects of distance shortening in the neighborhood of a singular fade subspace can be avoided by choosing a map which removes that singular fade subspace.

For SISO two-way relaying the singular fade subspaces are of the form {\scriptsize $\left( span \left(  \left[ \begin{array}{c}
 \Delta x_A \\ \Delta x_B
\end{array} \right]  \right)   \right)^{\dagger}  $ } =  $ span $ {\scriptsize $ \left( \left[ \begin{array}{c}
1 \\ - \frac{\Delta x_A}{\Delta x_B}
\end{array} \right] \right)  $ }, where $\Delta x_A = x_A - x_A'$ and $\Delta x_B = x_B - x_B'$ and $x_A,x_B,x_A',x_B' \in \mathcal{S}$. The ratio $-\frac{\Delta x_A}{\Delta x_B}$ determines all the singular fade subspaces and is called the \textit{singular fade state} \cite{NVR}, \cite{VvR_DSTC}. Procedures to obtain the maps $\mathcal{M}$ based on the channel fade state is given in \cite{APT1}, \cite{NVR}.

\section{Proposed Schemes}
Throughout the paper, a quasi static fading scenario is assumed with the channel state information is available only at the receivers (CSIR).
Let $\mathcal{S}$ denote the unit energy signal set used by the nodes A and B. A(B) wants to transmit $x_A(x_B) \in \mathcal{S}$  to B(A). Data transfer occurs in four phases: Two Multiple Access (MA) phases during which both the nodes A and B simultaneously transmit to the relay R followed by two Broadcast (BC) phases during which R transmits to the end nodes A and B. Hence the information rate in symbols per channel use in the proposed schemes is same as that of the schemes using DNF protocol in \cite{APT1}, \cite{NVR}.

In this paper we consider the following schemes:
\subsubsection*{Scheme-I} 
The nodes A and B are equipped with one antenna, while the relay R is equipped with two  antennas, but with only one RF chain. Hence at the relay during transmission, only one transmit antenna is active at a time. During reception any one of the antenna at R is active. 

\subsubsection*{Scheme-II} 
Here we consider the nodes A and B and the relay R are equipped with two antennas each, but with only one RF chain. Hence during the transmission from any node only one transmit antenna is active at a time. During reception any one of the antenna is active for all the three nodes. 

Compared to the schemes in \cite{APT1}, \cite{NVR}, \cite{VvR_DSTC}, the proposed schemes also use the same resources as the schemes uses only one RF chain.

\subsection{Scheme-I}
\label{APLNC_SM}
Let $h_{AR}$ and $h_{BR}$ denote the fade coefficients associated with A-R and B-R links respectively. Let $h_{R_iA}$ and $h_{R_iB}$ be the fade coefficients associated with R-A and R-B links respectively, where $i \in \{1,2\}$ denote the transmit antenna of the relay R. 

\subsubsection*{MA Phases}
The node A(B) wants to communicate two independent complex symbols $x_{A_1}(x_{B_1})$ and $x_{A_2}(x_{B_2})$ to B(A). During the $i$th MA phase $i \in \{1,2\}$, the node A(B) transmit $x_{A_i}(x_{B_i})$ to the relay R. The received signal at the relay R during the two MA phases is given by
\begin{eqnarray}
\nonumber
\bold{y_R} &  = & [y_{R_1} \ y_{R_2}] \\
 & = & \sqrt{E_s} \ [h_{AR} \ h_{BR}] \left[ \begin{array}{cc} x_{A_1} & x_{A_2} \\ x_{B_1} & x_{B_2} \\ \end{array} \right]  + [z_{R_1} \ z_{R_2}]
\label{eqn_yR}
\end{eqnarray}
where $E_s$ is the average transmission energy of the nodes. The complex symbol $y_{R_i}$, $i \in \{1,2\}$ denotes the received signal at R during the $i$th MA phase and $z_{R_i}$ denotes the additive noise and is distributed as $\mathcal{CN}(0,\sigma^2)$.

Let $(\hat{x}_{A_i}^R,\hat{x}_{B_i}^R) \in \mathcal{S}^2$ denote the Maximum Likelihood (ML) estimate of $(x_{A_i},x_{B_i})$ at R during the $i$th MA phase based on the received complex number $y_{R_i}$, i.e.,

{ \footnotesize
\begin{equation}
\nonumber
(\hat{x}_{A_i}^R,\hat{x}_{B_i}^R) = \arg\min_{(x_A,x_B) \in \mathcal{S}^2} |y_{R_i} - h_{AR}\sqrt{E_s}x_A - h_{BR}\sqrt{E_s}x_B |^2.
\end{equation} }

\subsubsection*{BC Phases}
Let $h_{BR}/h_{AR}$ be defined by $\gamma e^{j\theta}$. Depending on the value of $\gamma e^{j\theta}$, R chooses a many-to-one map $\mathcal{M}^{\gamma,\theta} : \mathcal{S}^2 \rightarrow \mathcal{S}' $, where $\mathcal{S}'$ is the unit energy signal set (of size between $M$ and $M^2$) having non-zero CPD used by the relay R during the BC phase. The map $\mathcal{M}^{\gamma,\theta}$ is chosen to avoid the harmful effects of distance shortening in the neighborhood of singular fade states and the procedure to find $\mathcal{M}^{\gamma,\theta}$ is given in \cite{NVR}. The choice of the map is indicated to the end nodes by overhead bits. To ensure that A(B) is able to decode B's(A's) message the map, $\mathcal{M}^{\gamma,\theta}$ should satisfy the exclusive law \cite{APT1}, i.e.,

{\footnotesize 
\begin{equation}
  \begin{aligned}
    \mathcal{M}^{\gamma,\theta}(x_A,x_B) & \neq \mathcal{M}^{\gamma,\theta}(x_A',x_B),&\mbox{for} \ x_A \neq x_A', \forall x_B \in \mathcal{S}, \\
\mathcal{M}^{\gamma,\theta}(x_A,x_B) & \neq \mathcal{M}^{\gamma,\theta}(x_A,x_B'),&\mbox{for} \ x_B \neq x_B', \forall x_A \in \mathcal{S}.
  \end{aligned} \bigg\}
  \label{exclusive_law}
\end{equation}  }

\begin{figure*}
\begin{align}
\nonumber
P\{E_{A_1}\} & = \mathbb{E} { \Bigg (} \frac{1}{M^2} \sum_{(x_{A_1},x_{B_1}) \in \mathcal{S}^2} { \bigg (} P_H\{E_{A_1}|\mathcal{M}^{\gamma,\theta}(\hat{x}_{A_1}^R,\hat{x}_{B_1}^R)\} =  \mathcal{M}^{\gamma,\theta} (x_{A_1},x_{B_1})\}P_H\{ \mathcal{M}^{\gamma,\theta}(\hat{x}_{A_1}^R,\hat{x}_{B_1}^R) =  \mathcal{M}^{\gamma,\theta} (x_{A_1},x_{B_1}) \} \\ \label{eqn:PH_EA}
& \hspace{1cm} +  P_H\{E_{A_1}|\mathcal{M}^{\gamma,\theta}(\hat{x}_{A_1}^R,\hat{x}_{B_1}^R)\} \neq  \mathcal{M}^{\gamma,\theta} (x_{A_1},x_{B_1})\}P_H\{ \mathcal{M}^{\gamma,\theta}(\hat{x}_{A_1}^R,\hat{x}_{B_1}^R) \neq  \mathcal{M}^{\gamma,\theta} (x_{A_1},x_{B_1}) \} { \bigg )} { \Bigg )}
\\
\nonumber
& \leq  \underbrace{\frac{1}{M^2} \sum_{(x_{A_1},x_{B_1}) \in \mathcal{S}^2} \mathbb{E} {\big (} P_H\{E_{A_1}|\mathcal{M}^{\gamma,\theta}(\hat{x}_{A_1}^R,\hat{x}_{B_1}^R) =  \mathcal{M}^{\gamma,\theta} (x_{A_1},x_{B_1})\} {\big )} }_{P^{A_1,BC}}  \\ \label{eqn:PH_EA_UB} 
& \hspace{5cm} +  \underbrace{\frac{1}{M^2} \sum_{(x_{A_1},x_{B_1}) \in \mathcal{S}^2} \mathbb{E} {\big (} P_H\{ \mathcal{M}^{\gamma,\theta}(\hat{x}_{A_1}^R,\hat{x}_{B_1}^R) \neq  \mathcal{M}^{\gamma,\theta} (x_{A_1},x_{B_1}) \} }_{P^{CE_1}} { \big )} 
\end{align}
\hrule
\end{figure*} 

\begin{figure*}
\begin{align}
\label{eqn:PH_BC}
P^{A_1,BC}  & =  \frac{1}{M^2} \sum_{(x_{A_1},x_{B_1}) \in \mathcal{S}^2} \mathop {\sum_{(x_{R_1}',x_{R_2},x_{R_2}') \in \mathcal{S'}^3}}_{x_{R_1}' \neq x_{R_1}} \mathbb{E} {\big (} P_H\{\hat{x}_{R_1} = x_{R_1}', \hat{x}_{R_2} = x_{R_2}',x_{R_2}|x_{R_1} =  \mathcal{M}^{\gamma,\theta} (x_{A_1},x_{B_1})\} {\big )}
\\
\label{eqn:PH_BC_UB}
& \leq \frac{1}{M^2} \sum_{(x_{A_1},x_{B_1}) \in \mathcal{S}^2} \mathop {\sum_{(x_{R_1}',x_{R_2},x_{R_2}') \in \mathcal{S'}^3}}_{x_{R_1}' \neq x_{R_1}} \mathbb{E} {\big (} P_H\{\hat{x}_{R_1} = x_{R_1}', \hat{x}_{R_2} = x_{R_2}'|x_{R_1} =  \mathcal{M}^{\gamma,\theta} (x_{A_1},x_{B_1}),x_{R_2}\} {\big )}
\end{align}
\hrule
\end{figure*}

Based on the ML estimates $(\hat{x}_{A_1}^R,\hat{x}_{B_1}^R)$ and $(\hat{x}_{A_2}^R,\hat{x}_{B_2}^R)$, the relay R need to transmit $x_{R_1} = \mathcal{M}^{\gamma,\theta}(\hat{x}_{A_1}^R,\hat{x}_{B_1}^R)$ and $x_{R_2} = \mathcal{M}^{\gamma,\theta}(\hat{x}_{A_2}^R,\hat{x}_{B_2}^R)$ during the broadcast phases. Let $\bold{h_{RA}} = [h_{R_1A} \ h_{R_2A}]$ and $\bold{h_{RB}} = [h_{R_1B} \ h_{R_2B}]$. The relay R transmits the two independent complex symbols $x_{R_1}$ and $x_{R_2}$ using a CIOD STBC given by $\bold{X_R} = \left[\begin{array}{cc}
\tilde{x}_{R_1} & 0 \\ 0 & \tilde{x}_{R_2}
\end{array} \right]$
in the two BC phases, where $\tilde{x}_{R_1}=x_{R_{1I}}+jx_{R_{2Q}}$ and $\tilde{x}_{R_2}=x_{R_{2I}}+jx_{R_{1Q}}$. Hence by using CIOD STBC, even though relay has two antennas, only one is active during one channel use. In each channel use, R has to switch the RF chain to the required antenna.  The received signals at the nodes A and B at the end of BC phases are respectively given by
\begin{eqnarray*}
\begin{array}{llll}
\bold{y_A} &= \ [y_{A_1} \ y_{A_2}] &=& \sqrt{E_s} \ \bold{h_{RA}X_R} + [z_{A_1} \ z_{A_2}] \\
\bold{y_B} &= \ [y_{B_1} \ y_{B_2}] &=& \sqrt{E_s} \ \bold{h_{RB}X_R} + [z_{B_1} \ z_{B_2}]
\end{array}
\end{eqnarray*}
where $z_{A_i}$ and $z_{B_i}$, $i\in \{1,2\}$ denote the additive noise and are distributed as $\mathcal{CN}(0,\sigma^2)$. Since A(B) knows its own messages, A(B) can decode $x_{B_i}(x_{A_i}), \ i\in \{1,2\} $ by decoding $x_{R_i}$.

\subsubsection*{Performance}
Let $\hat{x}_{B_1}^A (\hat{x}_{A_1}^B)$ and $\hat{x}_{B_2}^A (\hat{x}_{A_2}^B)$ denote the estimate of $x_{B_1}(x_{A_1})$ and $x_{B_2}(x_{A_2})$ respectively at the node A(B) at the end of BC phase. Due to symmetry, we can consider the SEP at one of the nodes. Let $E_{A_1}$ and $E_{A_2}$ represents error events $\{\hat{x}_{B_1}^A \neq x_{B_1}\}$ and $\{\hat{x}_{B_2}^A \neq x_{B_2}\}$ respectively. The average end-to-end SEP $P_S$, is given by 
\begin{equation}
P_S = \frac{1}{2} \left[ P\{E_{A_1}\}  + P\{E_{A_2}\} \right].
\label{eqn:PS}
\end{equation}
The symbols $x_{A_1}(x_{B_1})$ and $x_{A_2}(x_{B_2})$ are sent independently over two time slots by the node A(B), hence the symbols $x_{R_1}=\mathcal{M}^{\gamma,\theta}(\hat{x}_{A_1}^R,\hat{x}_{B_1}^R)$ and $x_{R_2}=\mathcal{M}^{\gamma,\theta}(\hat{x}_{A_2}^R,\hat{x}_{B_2}^R)$ to be transmitted in the BC phase are independent. Let $H=[h_{AR}~h_{BR}~h_{R_1A}~h_{R_2A}]$. The term $P\{E_{A_1}\}$ in (\ref{eqn:PS}) can be written as in (\ref{eqn:PH_EA}) and can be upper bounded by (\ref{eqn:PH_EA_UB}) as shown at the top of this page.

In (\ref{eqn:PH_EA}), $P_H\{ \mathcal{M}^{\gamma,\theta}(\hat{x}_{A_1}^R,\hat{x}_{B_1}^R)\} =  \mathcal{M}^{\gamma,\theta} (x_{A_1},x_{B_1}) \}$  and $P_H\{ \mathcal{M}^{\gamma,\theta}(\hat{x}_{A_1}^R,\hat{x}_{B_1}^R)\} \neq  \mathcal{M}^{\gamma,\theta} (x_{A_1},x_{B_1}) \}$ are the probabilities that R decodes to correct and wrong clusters respectively when the pair $(x_{A_1},x_{B_1})$ is transmitted by the end nodes, for a given realization of $H$. $P_H\{E_{A_1}|\mathcal{M}^{\gamma,\theta}(\hat{x}_{A_1}^R,\hat{x}_{B_1}^R)\} =  \mathcal{M}^{\gamma,\theta} (x_{A_1},x_{B_1})\}$ and $P\{E_{A_1}|\mathcal{M}^{\gamma,\theta}(\hat{x}_{A_1}^R,\hat{x}_{B_1}^R)\} \neq  \mathcal{M}^{\gamma,\theta} (x_{A_1},x_{B_1})\}$ are the probabilities of $E_{A_1}$ given that R decodes to correct and wrong clusters respectively, for a given $(x_{A_1},x_{B_1})$ pair and for a given $H$. The quantities $P^{A_1,BC}$ and $P^{CE_1}$ in (\ref{eqn:PH_EA_UB}), denote the probability of error event $E_{A_1}$ at node A at the end of the BC phase given that the relay decoded to the correct cluster at the end of MA phase and Cluster Error Probability (CEP) during the MA phase respectively.

Let $\hat{x}_{R_1}$ and $\hat{x}_{R_2}$ be the ML estimates $x_{R_1}$ and $x_{R_2}$ respectively at the node A at the end of BC phase. The error event $E_{A_1}$ is same as $\{\hat{x}_{R_1} \neq x_{R_1}\}$, given the relay decoded to correct cluster at the end of MA phase because the maps are required to satisfy the exclusive law (\ref{exclusive_law}). Since the relay decodes to the correct cluster, the symbol $x_{R_1}$ to be transmitted is $x_{R_1}=\mathcal{M}^{\gamma,\theta}(\hat{x}_{A_1}^R,\hat{x}_{B_1}^R) =  \mathcal{M}^{\gamma,\theta} (x_{A_1},x_{B_1})$. Now the term $P^{A_1,BC}$ can be written as (\ref{eqn:PH_BC}) and can be upper bounded by (\ref{eqn:PH_BC_UB}). The expression in (\ref{eqn:PH_BC_UB}) is the pairwise error probability (PEP) of a point-to-point $2\times1$ MIMO fading channel. From \cite{Khan_BSR},\cite{STC} PEP of a point-to-point $2\times1$ MIMO fading channel using a CIOD STBC has a diversity order of two and falls as $SNR^{-2}$.

\begin{figure*}
\begin{align}
\label{eqn:P_CE}
P^{CE_1} \leq 2P^{pp}(\mathcal{S}) + \frac{1}{M^2} \sum_{(x_{A_1},x_{B_1}) \in \mathcal{S}^2} \mathop{\sum_{{(x_{A_1}',x_{B_1}') \in \mathcal{S}^2}}}_{x_{A_1}' \neq x_{A_1}, x_{B_1}' \neq x_{B_1}} P^{CE}\{ (x_{A_1},x_{B_1}) \rightarrow (x_{A_1}',x_{B_1}')\}
\end{align}
\hrule
\end{figure*}

From \cite[Lemma 1]{VvR_perf}, the term $P^{CE_1}$ can be upper bounded by (\ref{eqn:P_CE}) shown at the top of the next page, where the second term corresponds the CEP associated with the removable singular fade states and $P^{pp}(\mathcal{S})$ denotes the average SEP of a point-to-point fading channel using the signal set $\mathcal{S}$ which falls with $SNR^{-1}$. For the ANC schemes, the diversity order associated with the removable singular fade states is two \cite[Th. 1]{VvR_perf}. The terms having diversity order of two in $P\{E_{A_1}\}$, i.e., CEP associated with removable singular fade states and broadcast error $P^{A_1,BC}$ can be neglected at high SNR. Hence, at high SNR  $P\{E_{A_1}\}$ can be upper bounded by $P\{E_{A_1}\} \leq 2P^{pp}(\mathcal{S})$.

Following a similar procedure, we can show that at high SNR the term $P\{E_{A_2}\}$ can also be upper bounded by $P\{E_{A_2}\} \leq 2P^{pp}(\mathcal{S})$. Hence from (\ref{eqn:PS}), the average end-to-end SEP can be upper bounded by

\begin{equation}
P_S \leq 2P^{pp}(\mathcal{S}).
\label{eqn:P_UB}
\end{equation}

Tight expressions for average SEP of a point-to-point fading channel are available in literature for commonly used signal sets and hence can be easily evaluated. Simulation results given in Section \ref{Sim_Results_Scheme_I} confirms that the upper bound in (\ref{eqn:P_UB}) is tight at high SNR.

\subsection{Scheme-II}
\label{PLNC_SM}
Extending the same idea of using the CIOD STBC's in BC phases to MA phases, we can get a diversity order of two for overall information transfer. Let $h_{A_iR}$ and $h_{B_iR}$ denote the fading coefficients between the $i$th transmit antenna of the nodes A and B respectively to the relay R, where $i\in \{1,2\}$. Let $h_{R_iA}$ and $h_{R_iB}$ be the fade coefficients between the $i$th transmit antenna of the relay R and the nodes A and B respectively, where $i \in \{1,2\}$. 

\subsubsection*{MA Phases}
Let $x_{A_1}(x_{B_1}) \in \mathcal{S}e^{j\theta}$ and $x_{A_2}(x_{B_2})  \in \mathcal{S}e^{j\theta}$ be the two independent complex symbols that the node A(B) want to communicate with the node B(A), where $\theta$ is the angle the signal set $\mathcal{S}$ is rotated so as to have non-zero CPD and maximum coding gain. Node A transmits two independent complex symbols $x_{A_1}$ and $x_{A_2}$ using a CIOD STBC matrix $\bold{X_A} = \left[\begin{array}{cc}
\tilde{x}_{A_1} & 0 \\ 0 & \tilde{x}_{A_2}
\end{array} \right]$in two time slots of the MA phases, where $\tilde{x}_{A_1}=x_{A_{1I}}+jx_{A_{2Q}}$ and $\tilde{x}_{A_2}=x_{A_{2I}}+jx_{A_{1Q}}$. Similarly node B transmits CIOD STBC matrix
$\bold{X_B} = \left[\begin{array}{cc}
\tilde{x}_{B_1} & 0 \\ 0 & \tilde{x}_{B_2} \end{array} \right]$, where $\tilde{x}_{B_1}=x_{B_{1I}}+jx_{B_{2Q}}$ and $\tilde{x}_{B_2}=x_{B_{2I}}+jx_{B_{1Q}}$.
Let $\bold{h} = [h_{A_1R} \ h_{A_2R} \ h_{B_1R} \ h_{B_2R}]$.
The received signal at R during the two MA phases can be written as 
\begin{eqnarray}
\nonumber
\bold{y_R} &  = & [y_{R_1} \ y_{R_2}] \\
 & = & \sqrt{E_s} \ \bold{h} \left[ \begin{array}{c} \bold{X_A} \\ \bold{X_B} \end{array} \right]  + [z_{R_1} \ z_{R_2}]
\label{eqn_yR_DSTC}
\end{eqnarray}
where $E_s$ is the average transmission energy of the nodes. $y_{R_i}$, $i \in \{1,2\}$ denotes the received signal at R during the $i$th MA phase and  $z_{R_i}$ denote the noise variable and distributed as $\mathcal{CN}(0,\sigma^2)$. The matrix,
\begin{equation}
\bold{C}(\bold{X_A},\bold{X_B}) = \left[ \begin{array}{c} \bold{X_A} \\ \bold{X_B} \end{array} \right] 
\label{PLNC_SM Matrix}
\end{equation}
is referred as a \textit{codeword matrix} for Scheme-II. \\

\noindent
\textit{B.1. With Fixed Network Coding}
\subsubsection*{BC Phases}
Let $(\hat{x}_{A_1}^R,\hat{x}_{A_2}^R,\hat{x}_{B_1}^R,\hat{x}_{B_2}^R)$ be the ML estimate of $(x_{A_1},x_{A_2},x_{B_1},x_{B_2})$ at the relay R. The relay R transmits the symbols $x_{R_1}=\hat{x}_{A_1}^R\oplus \hat{x}_{B_1}^R$ and $x_{R_2}=\hat{x}_{A_2}^R\oplus \hat{x}_{B_2}^R$ using a CIOD STBC matrix 
$\bold{X_R} = \left[\begin{array}{cc}
\tilde{x}_{R_1} & 0 \\ 0 & \tilde{x}_{R_2}
\end{array} \right]$, where $\tilde{x}_{R_1}=x_{R_{1I}}+jx_{R_{2Q}}$ and $\tilde{x}_{R_2}=x_{R_{2I}}+jx_{R_{1Q}}$ and $\oplus$ represents XOR map. XOR map is obtained by converting the complex symbols to bits, then doing bit-wise XOR operation, and converting back the bits to complex symbol. Let $\mathbf{h_{RA}} = [h_{R_1A} \ h_{R_2A}]$ and $\mathbf{h_{RB}} = [h_{R_1B} \ h_{R_2B}]$. The received signals at the nodes A and B at the end of the BC phases are respectively given by
\begin{eqnarray*}
\begin{array}{llll}
\bold{y_A} &= \ [y_{A_1} \ y_{A_2}] &=& \sqrt{E_s} \ \bold{h_{RA}X_R} + [z_{A_1} \ z_{A_2}] \\
\bold{y_B} &= \ [y_{B_1} \ y_{B_2}] &=& \sqrt{E_s} \ \bold{h_{RB}X_R} + [z_{B_1} \ z_{B_2}]
\end{array}
\end{eqnarray*}
where $z_{A_i}$ and $z_{B_i}$, $i\in \{1,2\}$ denote the additive noises and are distributed as $\mathcal{CN}(0,\sigma^2)$. Since A(B) knows its own messages and XOR map satisfies exclusive law, A(B) can decode $x_{B_i}(x_{A_i}), \ i\in \{1,2\} $ by decoding $x_{R_i}$.

Let $\bold{\Delta X_A} = \bold{X_A}-\bold{X_A'}$ and $\bold{\Delta X_B} = \bold{X_B}-\bold{X_B'}$. Let $\bold{C}(\bold{\Delta X_A},\bold{\Delta X_B}) = \bold{C}(\bold{X_A},\bold{X_B}) - \bold{C}(\bold{X_A'},\bold{X_B'}) $ denote the codeword difference matrix for Scheme-II, where $\bold{C}(\bold{X_A},\bold{X_B})$ is the codeword matrix defined in (\ref{PLNC_SM Matrix}). Since the CIOD STBC offers full diversity of two when the symbols are taken from a signal set having non-zero CPD, the codeword difference matrix $\bold{C}(\bold{\Delta X_A},\bold{\Delta X_B})$ is always full rank of 2, unless $\bold{\Delta X_A}=\bold{O}_2$ and $\bold{\Delta X_B}=\bold{O}_2$. Equivalently, all the pairwise error events $\bold{C}(\bold{X_A},\bold{X_B}) \rightarrow \bold{C}(\bold{X_A'},\bold{X_B'})$, have a diversity order of 2 \cite{STC}. Hence the diversity order for the SEP in the MA phases is two. In the BC phase, since CIOD STBC is transmitted from the relay to end nodes, the SEP at node A is of diversity order 2. Thus the average end to end SEP at node A will have a diversity order of 2. Simulation results are provided in Section \ref{Sim_Results_Scheme_II}.

The proposed scheme can offer a diversity order of two in contrast to the schemes in \cite{APT1}, \cite{NVR}, \cite{VvR_DSTC} which  provide a diversity order of one. The proposed scheme uses only one RF transmit chain in a time slot, hence without increasing the cost of implementation we can get a diversity gain. But this advantage for the proposed scheme come at a cost of increased decoding complexity at the relay R.

\subsubsection*{Decoding Complexity of Scheme-II at R}
The relay R need to jointly decode the symbols $(x_{A_1},x_{A_2},x_{B_1},x_{B_2})$ after the two MA phases. Generally the complexity of this joint ML decoding at R is $\mathcal{O}(M^4)$, where $M$ is the cardinality of the signal set $\mathcal{S}$. 
\begin{proposition}
\label{prop:decoding_complexity}
When conditional ML decoding \cite{BHV},\cite{Pavan_BSR}, is employed at R for the Scheme-II, the decoding complexity is $\mathcal{O}(M^3)$ when $\mathcal{S}$ is an  arbitrary signal set and is $\mathcal{O}(M^2\sqrt{M})$ when $\mathcal{S}$ is square QAM. 
\end{proposition}

\textit{Proof:}
For simplicity consider $E_s=1$.  Let $x_{A_i} = s_{A_i}e^{j\theta}$ and  $x_{B_i} = s_{B_i}e^{j\theta}$, where $s_{A_i} \in \mathcal{S}$ and $s_{B_i} \in \mathcal{S}$  for $i\in \{1,2\}$ and $\theta$ is the angle which the signal set $\mathcal{S}$ is rotated to have non-zero CPD and maximum coding gain. Let $\bold{\tilde{y}_R} = [y_{R_{1I}} \ y_{R_{1Q}} \ y_{R_{2I}} \ y_{R_{2Q}}]^T$, $\bold{\tilde{x}} = [x_{A_{1I}} \ x_{A_{1Q}} \ x_{A_{2I}} \ x_{A_{2Q}} \ x_{B_{1I}} \ x_{B_{1Q}} \ x_{B_{2I}} \ x_{B_{2Q}}]^T$,  $\bold{\tilde{z}_R} = [z_{R_{1I}} \ z_{R_{1Q}} \ z_{R_{2I}} \ z_{R_{2Q}}]^T$ and $\bold{\tilde{s}} = [s_{A_{1I}} \ s_{A_{1Q}} \ s_{A_{2I}} \ s_{A_{2Q}} \ s_{B_{1I}} \ s_{B_{1Q}} \ s_{B_{2I}} \ s_{B_{2Q}}]^T$. Let $\bold{V}_1 =\left[ \begin{array}{rr} \cos\theta & -\sin\theta \\ \sin\theta & \cos\theta
\end{array} \right]$ and let $\bold{V}=(\bold{I}_4 \otimes \bold{V}_1)$. We can see that $\tilde{\bold{x}} = \bold{V}\tilde{\bold{s}}$. The codeword matrix $\bold{C}(\bold{X_A},\bold{X_B})$ can be written as 
\begin{align}
\nonumber
{\raggedright \bold{C}(\bold{X_A},\bold{X_B}) = \sum_{i=1,2} \bold{W_{A}}_{iI}s_{A_{iI}} +  \bold{W_{A}}_{iQ}s_{A_{iQ}} }  \ \ \ \ \ \ \\
+ \bold{W_{B}}_{iI}s_{B_{iI}} + \bold{W_{B}}_{iQ}s_{B_{iQ}}
\end{align}
where $\bold{W_{A}}_{iI}$, $\bold{W_{A}}_{iQ}$, $\bold{W_{B}}_{iI}$ and $\bold{W_{B}}_{iQ}$ are referred to as the weight matrices.

The vector $\bold{\tilde{y}_R}$ can be written as $\bold{\tilde{y}_R}=
\bold{H_{eq}\tilde{x}} + \bold{\tilde{z}_R}$, where $\bold{H_{eq}} \in \mathbb{R}^{4\times8}$ whose entries are functions of $h_{A_1R}, \ h_{A_2R}, \ h_{B_1R}$ and $h_{B_2R}$ determined by the codeword matrix for Scheme-II. Now, $\bold{\tilde{y}_R}=\bold{\tilde{H}_{eq}\tilde{s}} + \bold{\tilde{z}_R}$, where $\bold{\tilde{H}_{eq}} = \bold{H_{eq}}\bold{V}$ . Using $\bold{QR}$ decomposition, $\bold{\tilde{H}_{eq}}$ can be decomposed into $\bold{\tilde{H}_{eq}}=\bold{QR}$, where $\bold{Q}\in \mathbb{R}^{4\times4}$ is a real orthogonal matrix and $\bold{R}\in \mathbb{R}^{4\times8}$ . The matrix $\bold{R}$ can be written as $[\bold{R_1} \ \bold{R_2}]$ where $\bold{R_1}$ is upper triangular. At R, the joint ML decoding metric is given by $||\bold{\tilde{y}_R}-\bold{\tilde{H}_{eq}\tilde{s}}||_2$ = $||\bold{Q^T\tilde{y}_R} - \bold{R\tilde{s}}||_2$ = $||\bold{y_R'} - \bold{R\tilde{s}}||_2$, where $\bold{y_R'} = \bold{Q^T\tilde{y}_R}$. 

It can be shown that the following pair of weight matrices are Hurwitz-Radon Orthogonal \footnote{Two complex matrices $\bold{M}_1$ and $\bold{M}_2$ are said to be Hurwitz-Radon orthogonal if $\bold{M}_1\bold{M}_2^H + \bold{M}_2\bold{M}_1^H$ = 0.} : $\{\bold{W_A}_{1I}, \bold{W_A}_{2I}\}$, $\{\bold{W_A}_{1I}, \bold{W_A}_{2Q}\}$, $\{\bold{W_A}_{1Q}, \bold{W_A}_{2I}\}$, $\{\bold{W_A}_{1Q}, \bold{W_A}_{2Q}\}$,  $\{\bold{W_B}_{1I}, \bold{W_B}_{2I}\}$, $\{\bold{W_B}_{1I}, \bold{W_B}_{2Q}\}$,  $\{\bold{W_B}_{1Q}, \bold{W_B}_{2I}\}$, $\{\bold{W_B}_{1Q}, \bold{W_B}_{2Q}\}$.
Let $\tilde{s}_i$ denotes the $i$th component of the vector $\tilde{\bold{s}}$.  From \cite[Th. 2]{Pavan_BSR}, the $i$th and $j$th columns of $\bold{\tilde{H}_{eq}}$ are orthogonal and hence the ($i,j$)th entry of $\bold{R}(i\leq j)$ is zero, if the weight matrices corresponding to the symbols $\tilde{s}_i$ and $\tilde{s}_j$  are Hurwitz-Radon orthogonal. Hence the structure of $\bold{R}$ will be of the form,
\[ \bold{R} = \left[ \begin{array}{cccccccc}
* & * & 0 & 0 & * & * & * & * \\
0 & * & 0 & 0 & * & * & * & * \\
0 & 0 & * & * & * & * & * & * \\
0 & 0 & 0 & * & * & * & * & * 
\end{array} \right] , \]
where $*$ denotes non-zero entries.

It can be seen from the matrix $\bold{R}$, that conditioning on the symbols $s_{B_1}$ and $s_{B_2}$ , the symbols $s_{A_1}$ and $s_{A_2}$ can be decoded independently \cite{Pavan_BSR}. Decoding $s_{A_1}$ and $s_{A_2}$ independently requires $2M$ computations and total number of choices for $s_{B_1}$ and $s_{B_2}$ is $M^2$. Hence the decoding requires $2M^3$ computations and hence the decoding complexity at R is $\mathcal{O}(M^3)$. For square QAM signal sets the complexity can be further reduced. Conditioning on $s_{B_1}$, $s_{B_2}$ and imaginary part of $s_{A_1}$, the real part of $s_{A_1}$ can be decoded independently. Similarly real part of $s_{A_2}$ can be decoded independently by conditioning on $s_{B_1}$, $s_{B_2}$ and imaginary part of $s_{A_2}$. Hence the total computations required is $2M^2\sqrt{M}$ and the complexity is $\mathcal{O}(M^2\sqrt{M})$ for square QAM signal set. \hfill $\blacksquare$ \\

\noindent
\textit{B.2. With Adaptive Network Coding}\\
In the previous subsection Fixed Network Coding (FNC) was used for the Scheme-II. But from Section \ref{SFC}, it is clear that there exists singular fade subspaces for Scheme-II. Hence by adaptively changing the maps based on the channel realization, we can further improve the end-to-end SEP performance. Let $\mathcal{V}$ be the fade space $[h_{A_1R} \ h_{A_2R} \ h_{B_1R} \ h_{B_2R}]$. During the BC phase for the ANC scheme, R chooses many to one map $\mathcal{M}^{\mathcal{V}} : \mathcal{S}^4 \rightarrow \mathcal{S'}$, where $\mathcal{S'}$ is a subset of $\mathbb{C}^2$. Based on the ML estimates $(\hat{x}_{A_1}^R,\hat{x}_{A_2}^R,\hat{x}_{B_1}^R,\hat{x}_{B_2}^R)$ at R, the relay R need to transmit the symbols $(x_{R_1},x_{R_2})$ = $\mathcal{M}^{\mathcal{V}}(\hat{x}_{A_1}^R,\hat{x}_{A_2}^R,\hat{x}_{B_1}^R,\hat{x}_{B_2}^R)$ during the BC phase. The maps can be obtained using the Latin squares approach in \cite{NVR}, \cite{VvR_graph}. The choice of map is indicated to the end nodes by the relay using overhead bits. The relay R transmits the complex symbols $x_{R_1}$ and $x_{R_2}$ using a CIOD STBC given by $\bold{X_R} = \left[\begin{array}{cc}
\tilde{x}_{R_1} & 0 \\ 0 & \tilde{x}_{R_2}
\end{array} \right]$
in the two BC phases, where $\tilde{x}_{R_1}=x_{R_{1I}}+jx_{R_{2Q}}$ and $\tilde{x}_{R_2}=x_{R_{2I}}+jx_{R_{1Q}}$. In order to decode the symbol transmitted by node A(B) at B(A), the map $\mathcal{M}^{\mathcal{V}}$ must satisfy the exclusive law. Simulation results given in Section \ref{Sim_Results_Scheme_II}, shows the advantage of using ANC instead of FNC at the relay.  The coding gain achieved by using ANC scheme over FNC scheme comes at a cost of increased decoding complexity at R mentioned in Proposition \ref{prop:decoding_complexity} and the efforts in finding the maps to be used at R for each channel realization \cite{NVR}.

\section{Simulation Results}
\label{Sim_Results}
In this section simulation results are presented for the proposed schemes when the end nodes use 4-QAM signal set and 8-PSK signal set having unit energy for both.

\begin{figure}[htbp]
\centering
\includegraphics[totalheight=2.5in,width=3.65in]{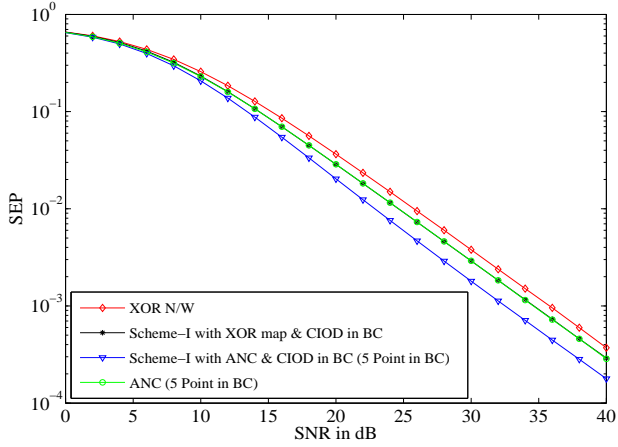}
\caption{SNR vs SEP plots for 4-QAM signal set for different schemes for Rayleigh fading scenario}	
\label{fig:Plots_4QAM_APLNC_SM}
\end{figure}

\begin{figure}[htbp]
\centering
\includegraphics[totalheight=2.5in,width=3.65in]{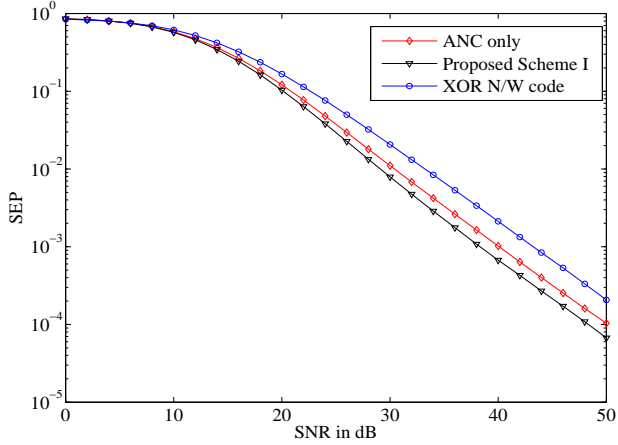}
\caption{SNR vs SEP plots for 8-PSK signal set for different schemes for Rayleigh fading scenario}	
\label{fig:Plots_8QAM_APLNC_SM}
\end{figure}

\subsection{Scheme-I}
\label{Sim_Results_Scheme_I}
The plots showing the coding gain achieved by Scheme-I compared to the schemes in \cite{APT1}, \cite{NVR} are given in Fig.\ref{fig:Plots_4QAM_APLNC_SM} and Fig.\ref{fig:Plots_8QAM_APLNC_SM} when the end nodes use 4-QAM signal set and 8-PSK signal set respectively for Rayleigh fading scenario, i.e, fade coefficients are distributed as $\mathcal{CN}(0,1)$. It can be seen from the plots that Scheme-I performs better than the ANC schemes mentioned in \cite{APT1} ,\cite{NVR} because the BC phase errors become less dominant as they have a diversity order of 2. Also if FNC is employed, proposed Scheme-I will perform better than the XOR network code \cite{APT1}.

\begin{figure}[htbp]
\centering
\includegraphics[totalheight=2.5in,width=3.65in]{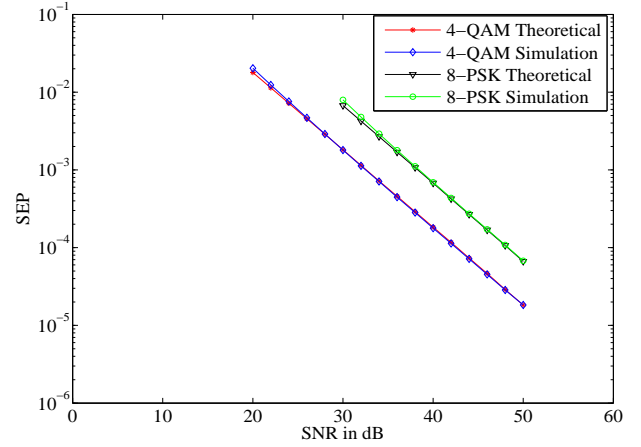}
\caption{Theoretical and simulation plots of SNR vs SEP for 4-QAM and 8-PSK}	
\label{fig:Performance_4QAM_8PSK_Sim_Theoretical}
\end{figure}

The high SNR upper bound on the average end-to-end SEP obtained in Section \ref{APLNC_SM} is evaluated and compared with the average end-to-end SEP obtained through simulations in Fig.\ref{fig:Performance_4QAM_8PSK_Sim_Theoretical} for 4-QAM and 8-PSK signal set for Rayleigh fading scenario. It can be seen from the plots that the bound obtained in Section \ref{APLNC_SM} is tight at high SNR. Note that this upper bound on average end-to-end SEP is valid only at high SNR. At low SNR it doesn't hold. 

\begin{figure}[htbp]
\centering
\includegraphics[totalheight=2.5in,width=3.65in]{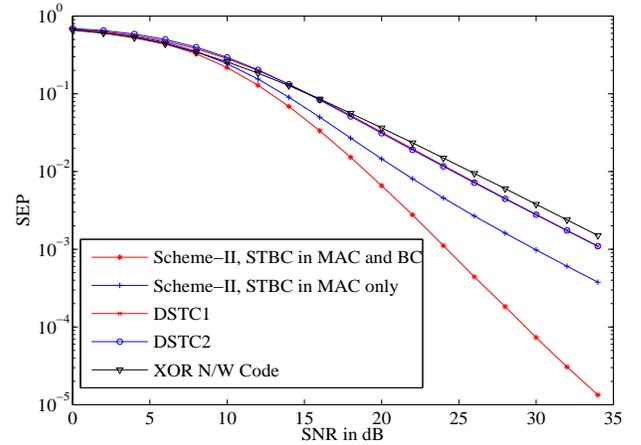}
\caption{SNR vs SEP plots for 4-QAM signal set for different schemes using XOR map in the BC phase}	
\label{fig:plots_4QAM_PLNC_SM}
\end{figure}

\subsection{Scheme-II}
\label{Sim_Results_Scheme_II}
\begin{figure}[htbp]
\centering
\includegraphics[totalheight=2.5in,width=3.65in]{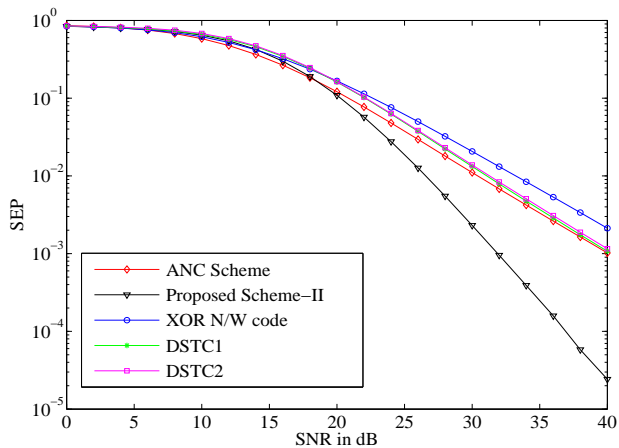}
\caption{SNR vs SEP plots for 8-PSK signal set for different Schemes}	
\label{fig:plots_8PSK_PLNC_SM}
\end{figure}

The plots showing the diversity gain for the proposed Scheme-II is given in Fig. \ref{fig:plots_4QAM_PLNC_SM} and Fig. \ref{fig:plots_8PSK_PLNC_SM}. SNR vs SEP plots for 4-QAM signal set for different Schemes using XOR network coding map at the relay R is given in Fig. \ref{fig:plots_4QAM_PLNC_SM}. DSTC1 and DSTC2 refers to the distributed space time codes in \cite{VvR_DSTC}, which uses two MA phases and two BC phases and uses XOR network map at R. From the plots, it can be seen that for the Scheme-II discussed in Section\ref{PLNC_SM}, diversity order of two can be achieved in contrast to other schemes. If CIOD STBC is used only in MA phases we will get a diversity order of one only as the broadcast errors of diversity one, determines the performance. But the performance in this case is better than other schemes in \cite{APT1}, \cite{NVR}, \cite{VvR_DSTC} because the errors in the multiple access phase is significantly reduced by using full diversity space time coding in MA phases. Fig. \ref{fig:plots_8PSK_PLNC_SM} shows the SNR vs SEP plots for 8-PSK signal set for Scheme-II and compares with XOR network coding,  ANC scheme and DSTC's.

\begin{figure}[htbp]
\centering
\includegraphics[totalheight=2.5in,width=3.65in]{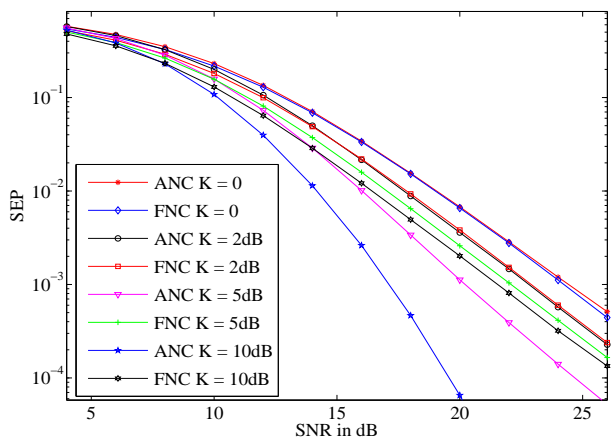}
\caption{SNR vs SEP plots for Scheme-II with ANC and FNC for different Rician Factors for 4-QAM signal set}	
\label{fig:plots_4QAM_ANC_FNC}
\end{figure}

Plots showing the advantage of using ANC scheme instead of FNC for transmission Scheme-II, is given in Fig. \ref{fig:plots_4QAM_ANC_FNC}. Plots are shown for 4-QAM signal set for Rayleigh fading scenario and Rician fading scenario with a \footnote{In a Rician fading channel with Rician fading factor K, the fade coefficient $X$ can be written as $\sqrt{\frac{K}{K+1}} + \frac{1}{\sqrt{K+1}}X_c$, where $X_c \sim \mathcal{CN}(0,1)$. A Rician fading channel with $K=0$ reduces to a Rayleigh fading channel.}Rician factor $K$. The maps are obtained using Latin squares approach mentioned in \cite{NVR}, \cite{VvR_graph}. Obtaining network coding maps by solving Latin squares was shown to be equivalent to proper vertex coloring of the singularity removal graph \cite{VvR_graph}. Using 4-QAM signal set for Scheme-II with ANC, the number of singular fade subspaces was found to be 804. The $16\times16$ Latin squares was solved using greedy algorithm by coloring the highest degree vertex first, mentioned in \cite{west}, as finding an optimal coloring is NP-complete. By using the greedy algorithm 16 to 24 symbols were required to complete the Latin squares for the removal of all the singular fade subspaces. Since in the BC phase CIOD STBC matrix conveys two complex symbols we choose a 5-point constellation. It can be seen from the plots that for $K=0$, i.e. Rayleigh fading, the average end-to-end SEP performance is worse for ANC scheme compared to FNC because of the usage 5-point constellation in BC phase. But ANC scheme performs better than FNC scheme when a significant line of sight component is present in the channel, as BC phase errors becomes less dominant compared to MAC phase errors.

\section{Discussion}
A transmission scheme based on physical layer network coding with space time coding using spatial modulation approach for two-way relaying scenario has been proposed. The proposed schemes is shown to perform better compared to ANC schemes and distributed space time codes for two-way relaying. For the proposed transmission Scheme-I, the average end-to-end SEP was shown to be upper bounded by twice the SEP of a point-to-point fading channel at high SNR. Diversity gain of two is achieved for the proposed transmission Scheme-II, with an increase in the decoding complexity at R. Removal of all the singular fade states for Scheme-II is shown to perform better than using fixed network coding. A direction for future work is to find the optimal coloring of the singularity removal graph obtained from the singularity removal constraints when CIOD's are used in the MA phase of relaying protocol and to complete the higher order partially filled Latin squares with optimum number of symbols.

\section*{Acknowledgment}
The authors would like to thank Vijayvaradharaj for his help and useful discussion on the subject of this paper.

%\vspace{-.2 cm}

\end{document}